\documentclass[letterpaper, 10 pt, conference]{ieeeconf}  

\author{Batool Ibrahim$^{1}$, Imad H. Elhajj$^{1}$, Daniel Asmar$^{2}$, Rawan El Hakim$^{1}$%
\thanks{$^{1}$Department of Electrical and Computer Engineering,
American University of Beirut, Beirut, Lebanon.}%
\thanks{$^{2}$Department of Mechanical Engineering,
American University of Beirut, Beirut, Lebanon.}%
}

\IEEEoverridecommandlockouts                              

\title{\LARGE \bf
Human-Robot Shared Control for Humanized End-Effector Teleoperation
}
\usepackage{float}
\usepackage{booktabs}
\usepackage{multirow}
\usepackage{pdflscape}
\usepackage{graphicx}
\usepackage{subcaption}
\usepackage{amsmath}
\usepackage{xcolor}
\begin{document}

\maketitle
\thispagestyle{empty}
\pagestyle{empty}


\begin{abstract}
Recent advances in robotics have enabled robots to operate in shared human environments, emphasizing the importance of effective human robot interaction HRI. Prior studies indicate that anthropomorphism, defined as the incorporation of human like features into robotic systems, facilitates more natural interaction and enhances both task performance and user experience. In robotic arm teleoperation, however, user controlled motions often deviate from human like kinematic characteristics due to intrinsic limitations of teleoperation systems. In this work, we propose a real time framework that generates human like end effector trajectories based on the two thirds power law of voluntary human hand movements, while preserving the operator’s intended control inputs. The proposed approach is validated through real world experiments conducted on a 6 degree of freedom Dobot CR10 robotic arm. Quantitative analysis demonstrates that the generated trajectories exhibit significantly stronger adherence to human like kinematic profiles compared to conventional teleoperation, with the estimated $\beta$ coefficient moving 39.7\% closer on average to the theoretical value of $-\tfrac{1}{3}$. Furthermore, the method achieves an approximate 34\% improvement in motion smoothness, measured by RMS torque rate reduction, with 80\% of evaluated motion patterns showing statistically significant improvements while maintaining comparable task completion times.
\end{abstract}

\section{INTRODUCTION}
With rapid advancements across domains such as healthcare, transportation, agriculture, and manufacturing \cite{niku2020introduction}, robots are increasingly deployed in tasks that involve interaction with humans \cite{sharkawy2021human}. These interactions may occur either remotely or in close physical proximity and range from passive observation, such as autonomous floor cleaning \cite{prabakaran2018floor}, to verbal communication, such as providing guidance in airports \cite{joosse2017guide}, and physical collaboration in shared manipulation tasks \cite{li2016adaptive}. As a result, Human Robot Interaction HRI has emerged as a major research area within robotics, attracting significant attention in recent years \cite{su2023recent}.

A central challenge in Human Robot Interaction HRI is improving human acceptance of robots in collaborative and interactive settings. Acceptance is closely associated with perceived trust, safety, and comfort \cite{firmino_de_souza2025_trust}, which are essential for effective human robot teamwork \cite{chen2018planning}. Empirical evidence indicates that robots exhibiting human like features are more readily accepted by users \cite{balcha2025review}, receive more positive HRI evaluations, and benefit from greater user tolerance to errors \cite{hri_explainations_2002}. Motivated by these findings, the robotics community increasingly integrates human inspired elements into HRI system design. Such efforts include replicating human form, as in humanoid robots \cite{ishiguro2001robovie}; human traits, such as deception or cheating behaviors \cite{esposito2025deception}; and human like motion characteristics \cite{gielniak2013generating}. This design approach is commonly referred to as robot anthropomorphism \cite{thomaz2013robotic}.

This paper investigates anthropomorphic robot motion derived from principles and strategies of human motor control, specifically how the central nervous system plans and executes movement through coordinated joints, muscles, and sensory feedback \cite{rosenbaum2009human}. Examples include goal directed reaching \cite{chatziparaschis2024adaptive}, grasping \cite{romano2011human}, and locomotion \cite{buschmann2017bioinspired}. Embedding human like motion characteristics in HRI enhances motion legibility, enabling users to better anticipate and interpret robot actions \cite{lichtenthaeler2016legibility}. This is especially critical in scenarios that rely on visual feedback and intuitive control, where predictable behavior supports trustworthy interaction \cite{khan2024hri_interfaces}. Teleoperation is a particularly challenging example of such scenarios.

Despite advances in automation, many systems still depend on human teleoperation via a Human Machine Interface HMI, particularly in robotic arm applications controlling the end effector. These systems are common in industrial tasks such as concrete spraying, polishing, and welding \cite{liu2024contact}, as well as creative tasks including robotic drawing and calligraphy \cite{chen2024autonomous}. However, teleoperated motions often lack human like kinematic properties due to indirect interaction through the HMI, absence of one to one kinematic mapping, task space mismatches, and communication delays \cite{lee2005bilateral}. Addressing these constraints motivates the central question of this work: how can human like motion be achieved under real time teleoperation constraints?

In this paper, we address this challenge by proposing a real time framework that integrates human motion characteristics into teleoperated robotic arms. Specifically, we anthropomorphize end effector trajectories to emulate the kinematic properties of human hand movements. The method is implemented through a shared control architecture in which both the human operator and the autonomous controller contribute to motion execution. This design preserves operator intent while enforcing human like kinematic structure. The adopted kinematic principle is the two thirds power law \cite{lacquaniti1983law}. Originally identified in the early 1980s, this law establishes a quantitative relationship between tangential velocity and trajectory curvature in human hand movements, whereby velocity decreases as path curvature increases.

The main contributions of this paper include the following:
\begin{itemize}
  \item We propose a real-time shared control framework for teleoperated robot arms that integrates operator commands with a kinematic model of human hand motion. This approach achieves human-like movements while preserving operator intent.  
  \item We demonstrate the effectiveness of the proposed approach through real-world experiments on a 6-DoF Dobot CR10 robot arm \cite{dobot_cr10}, showing that the executed end-effector motions exhibit more human-like characteristics compared to traditional operator-controlled motion.
  \item  We show that our approach not only achieves human-like motion during teleoperation, but also enhances motion smoothness compared to standard teleoperation, as quantified by torque based metrics.
  \item We provide our implementation as open-source code to facilitate further research in anthropomorphic teleoperation.
\end{itemize}

The remainder of this paper is structured as follows: Section II provides an overview of related works in the literature. Section III outlines the proposed system. Section IV presents a detailed analysis of the experimental setup and results. Finally, Section V offers conclusions and future directions.

\section{Related Work}
This section reviews existing literature relevant to our proposed framework. It focuses mainly on two principal areas: shared control in teleoperation and human-like motion generation in robotics. 

\subsection{Shared Control in Teleoperation}
Shared control was first introduced in 1978 as a control paradigm in which humans and machines simultaneously contribute to task execution. Owing to its capacity to combine human intent with robotic autonomy, shared control has been extensively investigated in the context of robotic arm teleoperation. Most existing studies focus on enabling autonomous obstacle avoidance during teleoperation, while the human operator specifies the desired motion through an interface such as a haptic device or joystick. For instance, Wang et al. \cite{wang2015shared} proposed a shared control framework that exploits joint redundancy and dimensionality reduction techniques to enhance obstacle avoidance. Rubagotti et al. \cite{rubagotti2023shared} addressed a similar objective using deep reinforcement learning, whereas Gottardi et al. \cite{gottardi2022shared} employed artificial potential fields. Comparable shared control strategies have also been applied to mobile robot teleoperation \cite{luo2019teleoperation} \cite{tian2021deep}. In all these works, the executed motions by the shared architecture are driven by task and environmental constraints, rather than by characteristics of human movement. In contrast, our framework realizes shared control by executing the operator trajectory commands while incorporating a kinematic model of the human hand movement. This ensures that the resulting motions demonstrate human-like characteristics.

More closely related to our work, Mower et al. \cite{mower2021skill} introduced a skill based shared control framework to support novice users in teleoperating robotic arms for industrial tasks such as spraying and polishing. Their approach estimates the intended human skill from operator inputs using optimization techniques and subsequently modifies the control signal to align with the inferred skill. However, the method is restricted to a finite set of predefined, task specific motion primitives, thereby limiting the execution of complex or previously unseen movements. In contrast, the proposed framework applies shared control at the level of fundamental human kinematic properties rather than task dependent skill primitives. This formulation enables users to execute arbitrary motions while ensuring that the resulting trajectories exhibit human like characteristics.

Tanwani et al. \cite{tanwani2017generative} addressed shared control through human intention estimation using learning from demonstration and probabilistic modeling techniques, including hidden semi Markov models. A similar approach was reported in \cite{xi2019robotic}. These methods offer adaptability to environmental variations and novel human motions by leveraging data driven representations. However, their performance depends on the availability of high quality demonstrations and periodic model updates to remain effective. In contrast, the proposed framework does not rely on prior learning data, thereby eliminating the need for demonstration collection or continuous retraining.

\subsection{Human-Like Motions in Robotics}
Human like motion generation in robotics constitutes a broad research area encompassing diverse forms of human movement embedded into robotic systems. Representative examples include locomotion patterns \cite{azevedo2004artificial}, reaching movements \cite{he2021anthropomorphic}, grasping behaviors \cite{zhu2021toward}, and human inspired trajectory planning strategies \cite{zhao2023human}.

Zhou et al. \cite{zhou2023humanlike} investigated robotic arm teleoperation using virtual reality trackers and proposed a human like inverse kinematics solver that integrates operator elbow constraints with conventional end effector constraints to generate anthropomorphic arm postures. Experimental results demonstrated increased user trust and reduced collision rates during teleoperation. In contrast, the present work focuses on embedding human like characteristics in the executed end effector trajectory rather than in the internal joint posture of the robotic arm.

Closely related to our work, Hicheur et al. \cite{rybarczyk2016effect} examined the effect of enforcing a human kinematic constraint in the teleoperation of mobile robots. They employed the same empirical relationship adopted in our framework, namely the two thirds power law \cite{lacquaniti1983law}, which relates motion velocity to trajectory curvature. In their study, the operator specified the desired movement direction via a touchscreen interface, after which the robot velocity was computed according to the geometric properties of the path and scaled based on the two thirds power law. Experimental results indicated that incorporating this law reduced task completion time, decreased collision frequency, and lowered user workload. While their approach applied the two thirds power law primarily for direct velocity scaling, our framework embeds it within a real time shared control architecture that supports continuous velocity based teleoperation. This enables the operator to retain full motion control rather than being limited to directional inputs. A comprehensive survey of human like motion generation methods is provided by Zhu et al. \cite{zhu2023human}.

Beyond control design, human motion characteristics have also been used for robotic system analysis and evaluation. Shafiei et al. \cite{shafiei2015using} applied the two thirds power law to segment robotic surgeons hand motions by analyzing affine velocity variations along the executed trajectory, with the aim of quantifying motion smoothness. Hugues et al. \cite{hugues2016determining} conducted a virtual reality user study to identify motion features that enhance human like perception. Subjective feedback indicated that human like end effector inertia reduced stress and increased motivation to collaborate, whereas velocity had a comparatively limited effect. However, the study was restricted to passive observation of a simulated industrial task, which may limit its applicability to physical robots and active shared control scenarios. Additional work has emphasized the role of human like velocity profiles in collaborative interaction. Maurice et al. \cite{maurice2017velocity} showed that enforcing two thirds power law velocity patterns reduced the force effort exerted by operators during direct end effector control and improved intuitive adaptation to robot motion. These findings underscore the relevance of incorporating biological kinematic laws into HRI frameworks.

\section{Proposed Architecture}
\begin{figure*}
    \centering
    \includegraphics[width=0.95\textwidth]{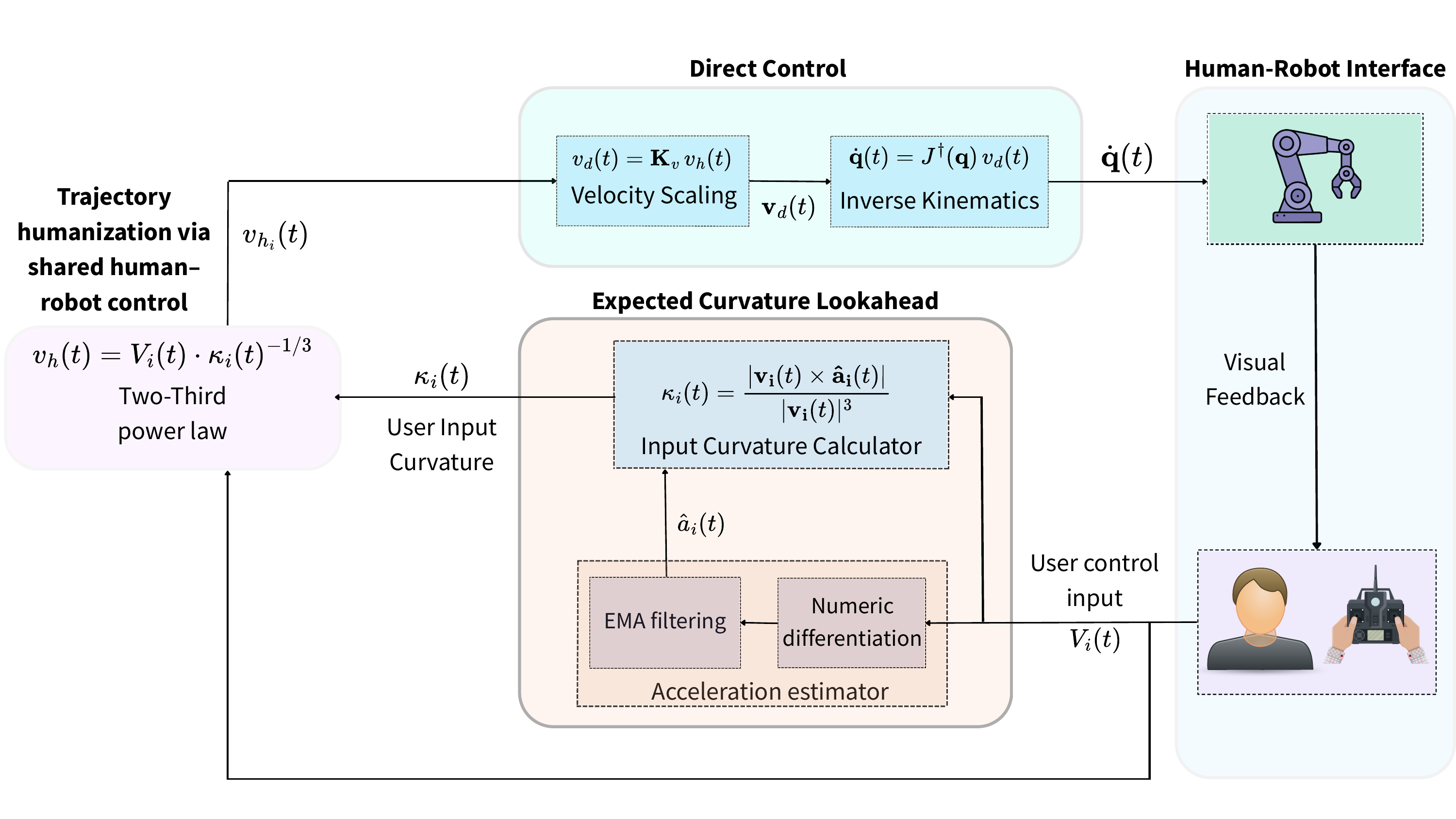}
    \caption{Block diagram of the shared human-robot control framework for human-like end-effector teleoperation. User input velocity $\mathbf{V}_i(t)$ is used to estimate the expected curvature $\hat{C}_i(t)$ of the desired motion. The velocity signal is then adjusted according to the two-thirds power law, producing humanized commands $\mathbf{v}_h(t)$ that preserve natural human motion characteristics.}
    \label{fig:proposed-model}
\end{figure*}

The proposed system architecture is illustrated in Fig. \ref{fig:proposed-model}. The operator commands the robot arm end effector through a joystick while relying on real time visual feedback to continuously adjust the input signals. Unlike conventional teleoperation, where operator velocity commands are directly executed, the proposed framework modulates these commands through a shared control scheme that embeds human hand kinematic behavior governed by the two thirds power law. Consequently, the control signals transmitted to the robot arm are designed to preserve the user intended motion while enforcing human like velocity profiles. The individual components of the system are detailed in the following section.

\subsection{Human-like Motion}
\subsubsection{\textbf{The Two-third power law}}
\label{2}
The two-thirds power law is expressed by the following mathematical relation:
\begin{equation}
\label{3}
v(t) = k \cdot C(t)^{-1/3}
\end{equation}
where $v(t)$ is the tangential velocity, $C(t)$ is the trajectory curvature, and $k$ is a positive constant that scales the overall speed of motion.  This relationship shows that tangential velocity is inversely proportional to the cube root of curvature. In other words, velocity decreases with sharper trajectories (higher curvature) and increases with smoother movements (lower curvature). The law applies particularly to planar motions, as it was formulated for voluntary hand movements such as writing and drawing.

\subsubsection{\textbf{Curvature Lookahead}}
\label{5}
The curvature of a path is formally defined as the magnitude of the rate of change of its unit tangent vector with respect to arc length. It is expressed as follows: 
\begin{equation}
C = \left\| \frac{d\mathbf{T}}{ds} \right\|,
\end{equation}
where $\mathbf{T}$ denotes the unit tangent vector and $s$ represents the arc length.  For time-parameterized trajectories, curvature can equivalently be expressed as:
\begin{equation}
C(t) = \frac{\|\mathbf{v}(t) \times \mathbf{a}(t)\|}{\|\mathbf{v}(t)\|^{3}},
\end{equation}
where $\mathbf{v}(t)$ and $\mathbf{a}(t)$ denote the velocity and acceleration vectors of a trajectory, respectively. This formulation of curvature is adopted in the proposed framework because the teleoperation scheme uses velocity vectors as the primary control input. As a result, curvature can be computed directly from the available kinematic quantities within the control pipeline, eliminating the need for explicit geometric path analysis.

The curvature lookahead block receives the user commands at time $t$ as input. These commands represent the operator desired end effector motion; however, due to teleoperation constraints, they do not inherently exhibit human like kinematic characteristics. Assuming planar motion in the $x$–$y$ plane, the input velocity vectors are expressed as follows:
\begin{equation}
\mathbf{V_i}(t) = [V_x(t), V_y(t)]
\end{equation}
These velocity vectors are used to estimate the user input acceleration through numerical differentiation:
\begin{equation}
\mathbf{a_i}(t) = \frac{\mathbf{V_i}(t) - \mathbf{V_i}(t-\Delta t)}{\Delta t}
\end{equation}
Exponential Moving Average (EMA) filtering is then applied to reduce the effect of noise in the differentiation process:
\begin{equation}
\hat{\mathbf{a}}_i(t) = \alpha. \mathbf{a_i}(t) + (1-\alpha).\hat{\mathbf{a}}_i(t-1),
\end{equation}
where $\alpha$ is the smoothing factor. Accordingly, the expected curvature of the user motion is computed using the filtered acceleration $\hat{\mathbf{a}}_i(t)(t)$ and the input velocity $\mathbf{V_i}(t)$:
\begin{equation}
\hat{C}_i(t) = \frac{||\mathbf{V_i}(t) \times \hat{\mathbf{a}}_i(t)||}{||\mathbf{V_i}(t)||^3}
\end{equation}

This predicts the curvature of the user motion prior to execution. The subsequent shared control module then integrates this predicted curvature with the user commands to modify the control signal according to the two-third power law relationship. 

\subsubsection{\textbf{Shared Control}}
\label{1}
The shared control component constitutes the core of human like motion generation in the proposed architecture. It receives as inputs the operator direct velocity commands and the predicted curvature provided by the curvature lookahead module. Based on these quantities, it enforces the kinematic characteristics of human hand motion to produce anthropomorphic control signals. Through this mechanism, both the human operator and the autonomous controller contribute to the final executed motion.

As discussed in Section \ref{2}, the two-thirds power law is expressed by \eqref{3}. We incorporate this law into the shared control framework by proposing the humanized velocity command as:
\begin{equation}
\label{4}
\mathbf{v}_h(t) = ||\mathbf{V}_i(t)|| \cdot \hat{C}_i(t)^{-1/3}
\end{equation}

More specifically, we propose replacing the gain factor $K$ with the instantaneous magnitude of the user input velocity, $|\mathbf{V}_i(t)|$. In the original formulation \cite{lacquanti2019global}, the gain factor scales the overall motion speed to reflect human motor planning characteristics. In contrast, within our teleoperation framework, the human operator remains actively engaged in the control loop and continuously specifies desired velocities through the interface. Therefore, the magnitude of the user input velocity vector, $|\mathbf{V}_i(t)|$, inherently captures the operator real time motor planning decisions and encodes the intended motion speed. By substituting $K$ with this quantity, the proposed method embeds the human like kinematic constraints imposed by the two thirds power law while fully preserving the operator real time control authority.

\subsection{Robot Arm Direct Motion Control}
Finally, the humanized control signal is forwarded to the robot arm control module for execution. The control module first applies velocity scaling, as the incoming velocity commands are typically normalized. This yields the final desired velocity command:
\begin{equation}
\mathbf{V}_d(t) = K_v \mathbf{v}_h(t),
\label{eq:desired_velocity}
\end{equation}
where $K_v$ presents the velocity scaling factor determined according to the robot kinematic and hardware specifications.

Since the desired task involves teleoperation of the robot end-effector, the velocity commands are translated into task-space. The robot joint configuration is represented by the vector $\mathbf{q}(t)$, and the end-effector position $\mathbf{p}(t)$.
The end-effector linear velocity is related to joint velocities through the Jacobian matrix:
\begin{equation}
\dot{\mathbf{p}}(t) = J\big(\mathbf{q}(t)\big) \cdot \dot{\mathbf{q}}(t)
\label{eq:forward_kinematics}
\end{equation}
To achieve the desired end-effector velocity, the corresponding joint velocities are computed as follows:
\begin{equation}
\dot{\mathbf{q}}(t) = J^\dagger\big(\mathbf{q}(t)\big)\,\mathbf{V}_d(t),
\label{eq:inverse_kinematics}
\end{equation}
where $J^\dagger$ denotes the damped pseudoinverse of the Jacobian matrix, defined as
\begin{equation}
J^\dagger = J^T \left( J J^T + \lambda^2 I \right)^{-1},
\label{eq:damped_pseudoinverse}
\end{equation}
with $\lambda$ representing the damping factor.

\section{Experiments and Results discussion}
\subsection{Implementation Details}
\begin{figure}[htbp]
    \centering
    \includegraphics[width=\linewidth]{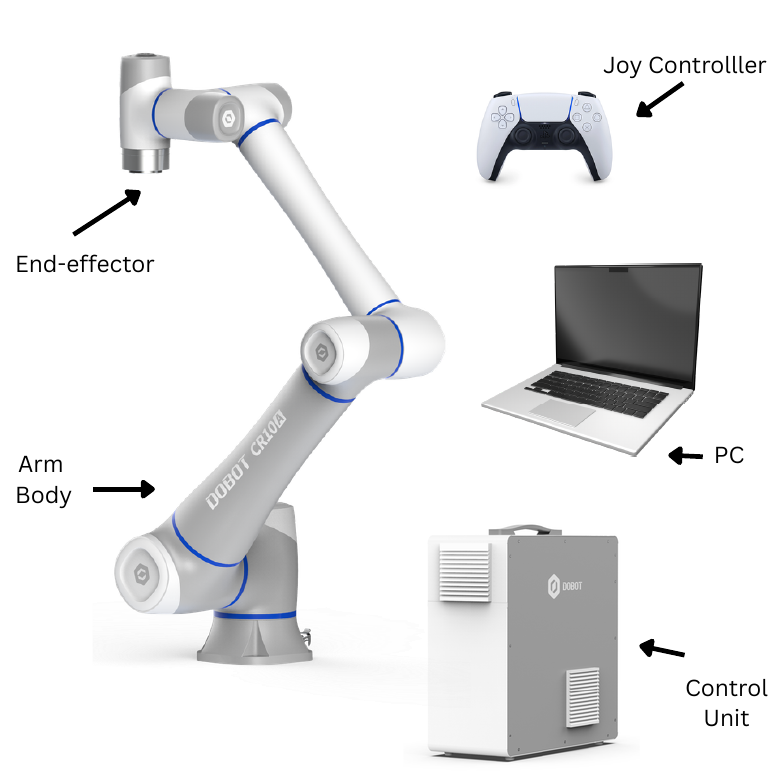}
    \caption{Experimental platform consisting of a Dobot CR10 robotic arm, joystick controller for user input, a control unit, and a PC running the proposed shared-control teleoperation framework.}
    \label{fig:dobot_cr10}
\end{figure}

To validate the proposed framework, experimental evaluation was conducted on a 6 DoF Dobot CR10 collaborative robotic arm. The experimental setup is presented in Fig.\ref{fig:dobot_cr10}. The control pipeline was implemented using the Robot Operating System ROS \cite{ros}, which provided the communication and control infrastructure. The human machine interface consisted of a joystick generating velocity commands. End effector control was implemented using MoveIt for motion planning and kinematic computations, together with MoveIt Servo, which enables smooth real time teleoperation through velocity based commands. The complete implementation is publicly available in our GitHub repository.

\subsection{Experimental Details}
The experimental protocol consisted of the following steps. First, a human operator teleoperated the robot end effector using a joystick based human machine interface, while the issued velocity commands were recorded. The recorded command sequences were subsequently replayed under two conditions: (1) conventional direct teleoperation and (2) the proposed shared control framework incorporating human like kinematic characteristics. Each trajectory was executed three times per condition to account for stochastic variations and to support statistical reliability.

Although the framework is designed for real time teleoperation, command replay was adopted to ensure identical input trajectories in both scenarios. This procedure eliminates variability introduced by human input and ensures a controlled, fair comparison. In addition, user commands are recorded with timestamps during real-time operation. During motion replay, the system reproduces the exact temporal sequence, behaving identically to live teleoperation.

The evaluated motions included planar trajectories commonly studied in the two-thirds power law literature: ellipses, figure-eight (infinity) patterns, sinusoidal trajectories, and arbitrary user-defined paths.
Throughout execution, all relevant sensor data were recorded, including joint positions, joint velocities, end-effector trajectories, and joint torques. Performance was evaluated by comparing the averaged metrics across three repetitions for each trajectory under the two experimental conditions.

\subsection{Metrics Used}
\label{10}
We consider the following metrics to quantitatively assess the performance of our proposed system compared to the traditional approach:

\subsubsection{Task Performance Metrics}
Completion Time ($T_c$), the duration required to complete the specified motion trajectory, measured from motion start to motion end.
    \begin{equation}
        T_c = t_{\text{end}} - t_{\text{start}}
    \end{equation}
\subsubsection{Human-Likeness Metrics}
\label{9}
is quantified through the analysis of the $\beta$ coefficient in the curvature--velocity power-law relationship:
\begin{equation}
v = K C^{\beta},
\end{equation}
For motion strictly following the two-thirds power law, the theoretical exponent satisfies $\beta = -\tfrac{1}{3}$. The coefficient $\beta$ is estimated via linear regression in log--log space, as performed in \cite{fraser2025biological}. 
Taking the logarithm of both sides yields
\begin{equation}
\log v = \log K + \beta \log C,
\end{equation}
which allows $\beta$ to be obtained from the slope of the fitted linear model. Human-likeness was therefore assessed by measuring the deviation of the estimated $\beta$ from the theoretical value $-\tfrac{1}{3}$.
\subsubsection{Motion Smoothness Metrics}
Based on torque-rate analysis. This is because shifts in joint torque are closely linked to how much the acceleration (or jerk) changes at each joint.

\begin{itemize}
    \item \textbf{Peak Torque Rate ($\Sigma$ Peak $|d\tau/dt|$):} The maximum instantaneous rate of torque change across all joints, measured in Nm/s.
    \begin{equation}
        \Sigma \text{Peak} \left|\frac{d\tau}{dt}\right| = \max_{t \in [t_{\text{start}}, t_{\text{end}}]} \sum_{j=1}^{6} \left|\frac{d\tau_j}{dt}\right|
    \end{equation}
    where $\tau_j$ is the torque of joint $j$. Lower values indicate smoother motion with reduced peak acceleration changes.

    \item \textbf{RMS Torque Rate ($\Sigma$ RMS $d\tau/dt$):} The root mean square of torque rate across all joints.
    \begin{equation}
        \Sigma \text{RMS} \frac{d\tau}{dt} = \sqrt{\frac{1}{N} \sum_{i=1}^{N} \left(\sum_{j=1}^{6} \frac{d\tau_j}{dt}\right)^2}
    \end{equation}
    where $N$ is the number of samples during motion. Lower RMS values indicate smoother motion throughout the trajectory.
\end{itemize}

Note that we adopted traditional teleoperation as a baseline because, to the best of our knowledge, the implementation of human-like motion in real-time teleoperation has not been previously explored in the literature. Therefore, there are no directly comparable state-of-the-art frameworks. Although alternative human motion models within the scope of the two-thirds power law are available, substituting the two-thirds power law with these models would primarily highlight differences between motion models rather than the effect of humanized teleoperation.

\begin{figure}[H]
    \centering
    \begin{subfigure}{\linewidth}
        \centering
        \includegraphics[width=0.8\linewidth]{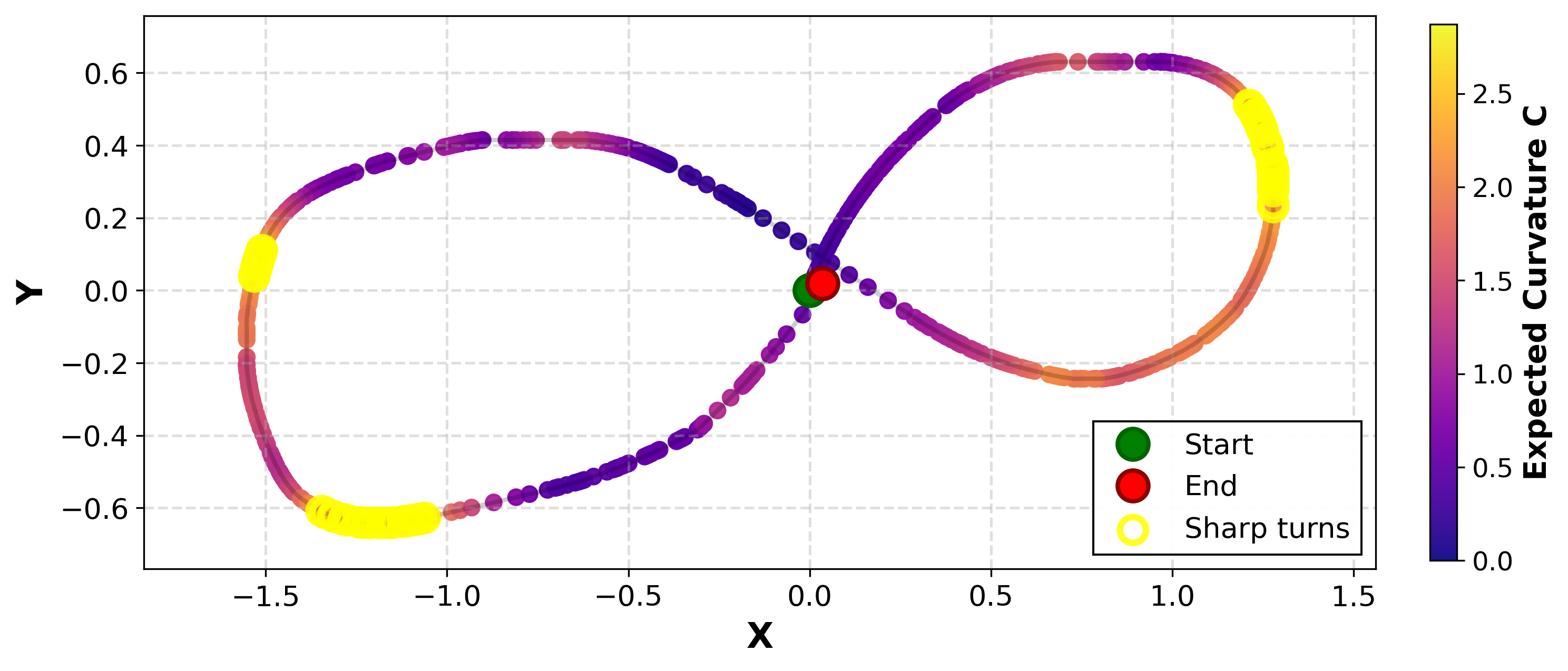}
        \caption{Trajectory in command space colored by expected curvature C.
        }
        \label{fig:trajectory15}
    \end{subfigure}
    
    \vspace{0.5em}
    
    \begin{subfigure}{\linewidth}
        \centering
        \includegraphics[width=0.8\linewidth]{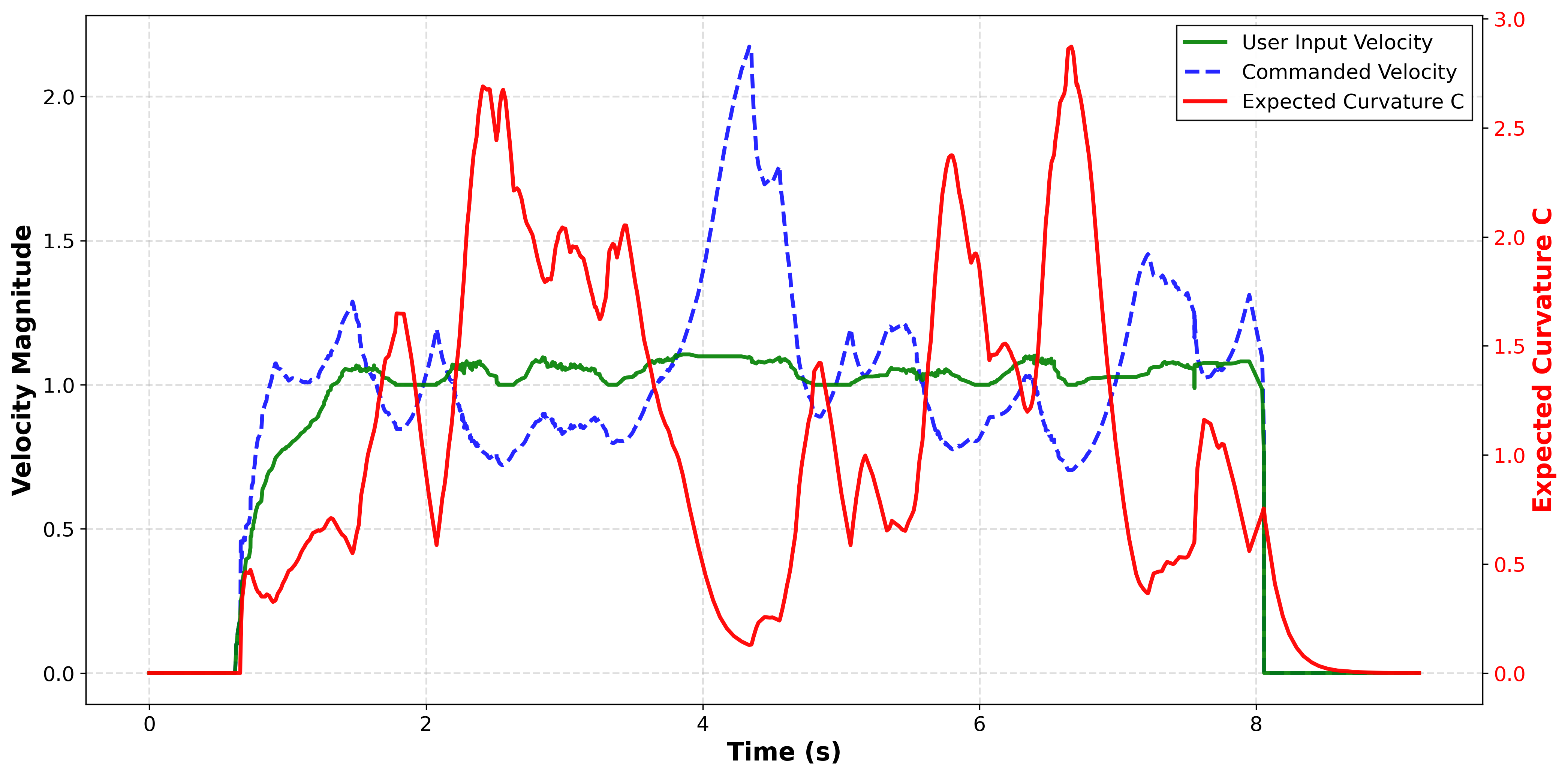}
        \caption{User input velocity (green), commanded velocity (blue, dashed), and expected curvature C (red, right axis) versus time.
        }
        \label{fig:curvature_velocity15}
    \end{subfigure}
    
    \caption{Trajectory analysis and velocity modulation.}
    \label{fig:dataset15}
\end{figure}

\begin{figure}[H]
    \centering
    
    \begin{subfigure}{\linewidth}
        \centering
        \includegraphics[width=0.8\linewidth]{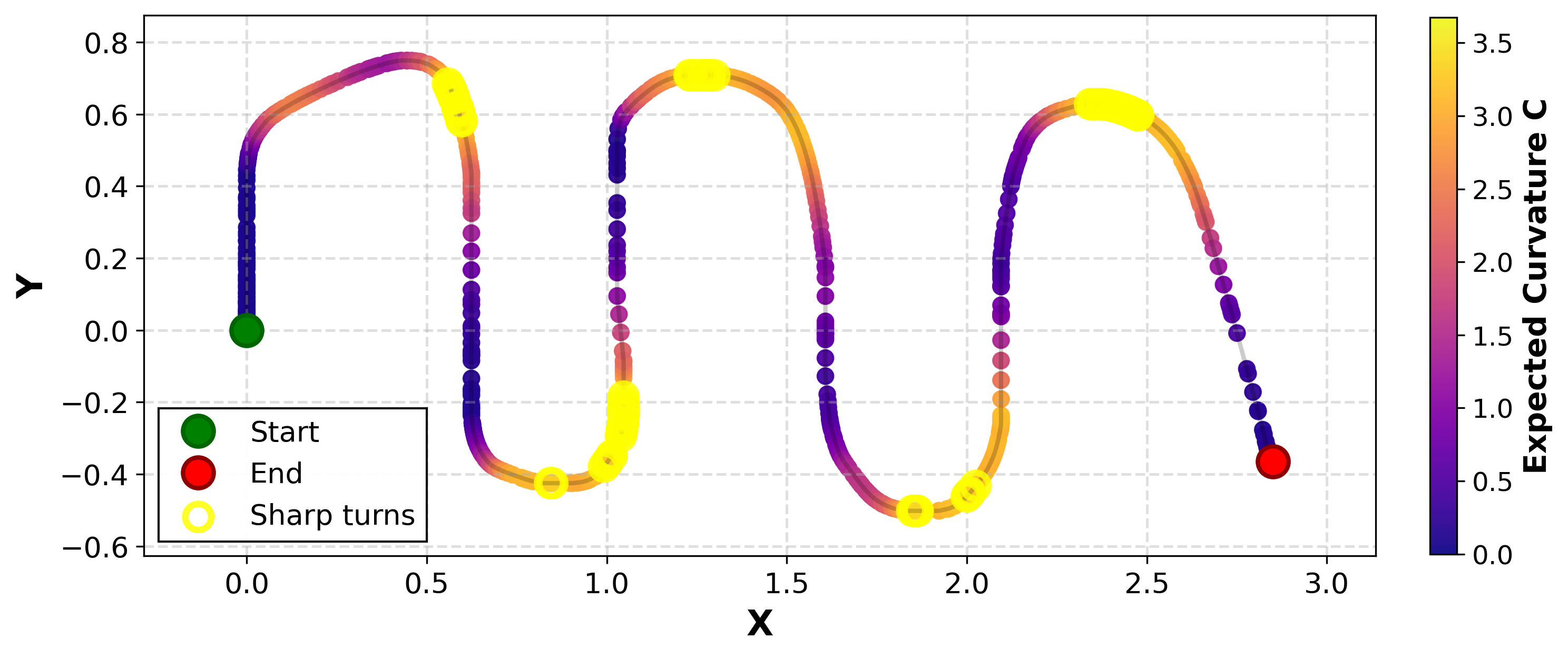}
        \caption{Trajectory in command space colored by expected curvature C. 
        }
        \label{fig:trajectory11}
    \end{subfigure}
    
    \vspace{0.5em}
    
    \begin{subfigure}{\linewidth}
        \centering
        \includegraphics[width=0.8\linewidth]{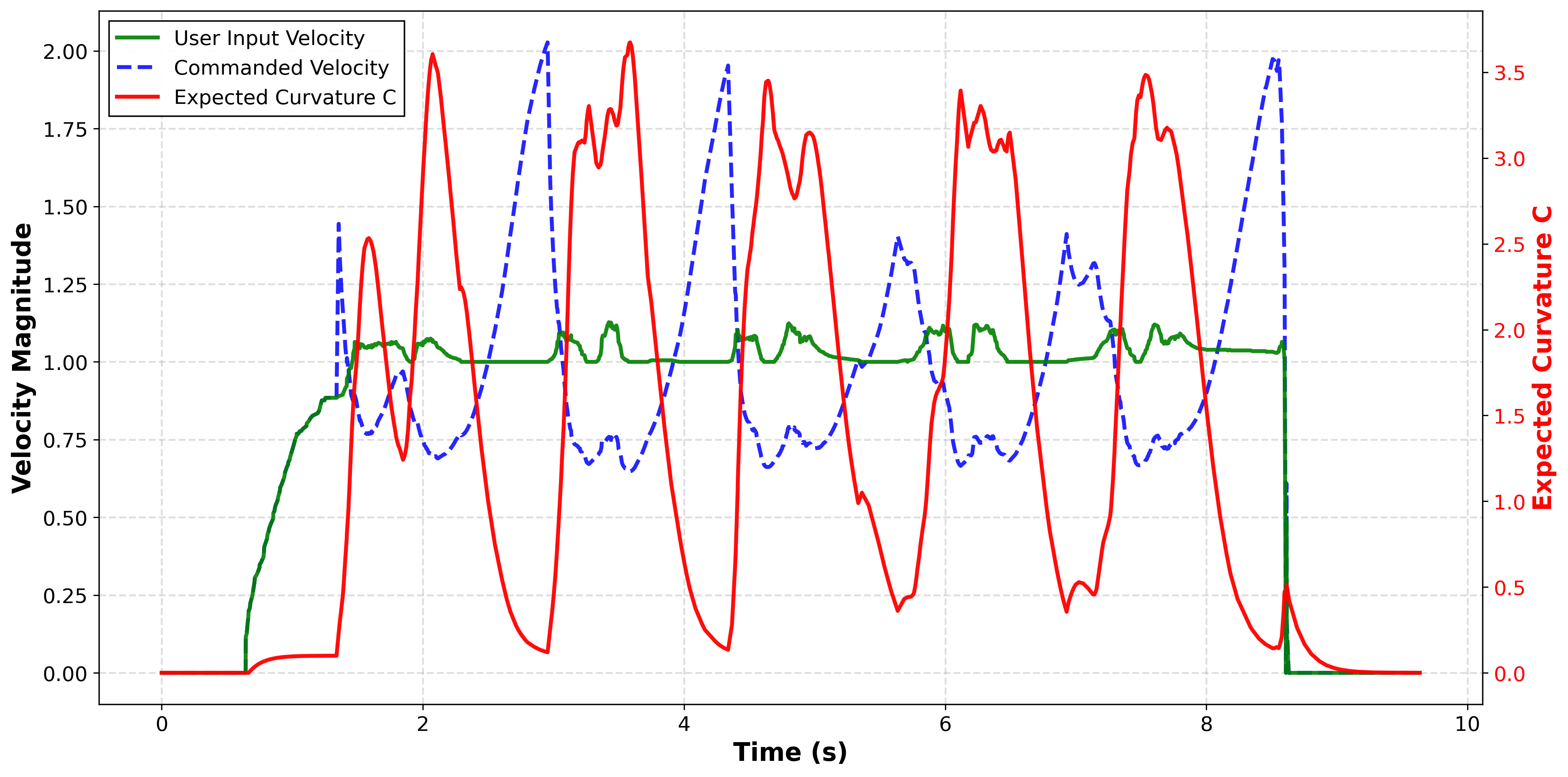}
        \caption{User input velocity (green), commanded velocity (blue, dashed), and expected curvature C (red, right axis) versus time.
        }
        \label{fig:curvature_velocity11}
    \end{subfigure}
    
    \caption{Trajectory and velocity analysis of one of our tests.}
    \label{fig:dataset11}
\end{figure}

\subsection{Results}
\subsubsection{Human-Likeness}
First, we investigate the input control signals before and after the humanization process within our proposed shared-control architecture. For clarity of discussion, two representative executed motions are considered. Figures \ref{fig:dataset15}(a) and \ref{fig:dataset11}(a) present the executed trajectories in command space, colored according to the expected curvature predicted by the lookahead module defined in Section~\ref{5}. As observed in the figures, predicted curvature values increase along trajectory segments corresponding to sharp directional changes and decrease along near-linear segments. This indicates that the curvature lookahead mechanism, operating on the user input commands, successfully anticipates upcoming geometric variations rather than reacting instantaneously. 

Moreover, Figures \ref{fig:dataset15}(b) and \ref{fig:dataset11}(b) present a comparison between the final commanded velocities and the direct user input prior to humanization, analyzed with respect to curvature variations. As shown, the user input velocity maintains an approximately constant magnitude when no human like modulation is applied. In contrast, the velocity generated by the proposed framework adheres to the two thirds power law, demonstrating an inverse relationship with curvature: velocity increases as curvature decreases and decreases as curvature increases.

\begin{table*}[t]
\centering
\small
\setlength{\tabcolsep}{2.5pt}
\caption{Power-law exponent ($\beta$) comparison across all motion patterns. The biological 1/3 power law predicts $\beta \approx -0.333$ for human-like motion. Bold indicates improvements (closer to human-like). $\Delta$\% represents percentage reduction in deviation from the ideal $\beta = -0.333$. Values shown as mean $\pm$ std across trials.}
\label{tab:powerlaw_beta}
\begin{tabular}{llcccc}
\toprule
\multirow{2}{*}{\textbf{Motion}} & \multirow{2}{*}{\textbf{Index}}
& \textbf{Traditional}
& \textbf{Proposed}
& \textbf{Deviation} & \textbf{More} \\
\cmidrule(lr){3-4}
& & $\beta$ (mean $\pm$ std) & $\beta$ (mean $\pm$ std) & $\Delta$ (\%) & \textbf{Human-like} \\
\midrule
Arbitrary & 1 & $-0.495 \pm 0.108$ & $-0.457 \pm 0.069$ & \textbf{+23.2\%} & \textbf{Proposed} \\
\midrule
\multirow{3}{*}{Ellipse} & 1 & $-0.259 \pm 0.009$ & $-0.309 \pm 0.013$ & \textbf{+67.8\%} & \textbf{Proposed} \\
& 2 & $-0.612 \pm 0.036$ & $-0.617 \pm 0.064$ & -2.1\% & Traditional \\
& 3 & $-0.505 \pm 0.132$ & $-0.416 \pm 0.049$ & \textbf{+51.9\%} & \textbf{Proposed} \\
\midrule
\multirow{3}{*}{Infinity} & 1 & $-0.585 \pm 0.285$ & $-0.402 \pm 0.026$ & \textbf{+72.9\%} & \textbf{Proposed} \\
& 2 & $-0.530 \pm 0.065$ & $-0.457 \pm 0.045$ & \textbf{+36.9\%} & \textbf{Proposed} \\
& 3 & $-0.383 \pm 0.074$ & $-0.338 \pm 0.078$ & \textbf{+90.5\%} & \textbf{Proposed} \\
\midrule
\multirow{3}{*}{Sinusoidal} & 1 & $-0.297 \pm 0.113$ & $-0.288 \pm 0.043$ & -26.9\% & Traditional \\
& 2 & $-0.693 \pm 0.106$ & $-0.545 \pm 0.146$ & \textbf{+41.1\%} & \textbf{Proposed} \\
& 3 & $-0.296 \pm 0.139$ & $-0.328 \pm 0.032$ & \textbf{+85.2\%} & \textbf{Proposed} \\
\midrule
\multicolumn{6}{l}{\small Reference: $\beta = -0.333$ (1/3 power law for biological motion)} \\
\multicolumn{6}{l}{\small Overall: Proposed closer to human-like in 8/10 patterns (80\%), average improvement: 39.7\%} \\
\bottomrule
\end{tabular}
\end{table*}


Table \ref{tab:powerlaw_beta} summarizes the quantitative results of the human likeness evaluation for each tested motion, as described in Section \ref{9}. All motions show a reduced deviation of the estimated $\beta$ coefficient from the theoretical value $-\tfrac{1}{3}$, indicating increased adherence to human like kinematic behavior. The average improvement across all trials was $+39.7\%$, with 8 out of 10 motion patterns demonstrating measurable enhancement. Although the estimated exponents did not converge exactly to the theoretical value, this result is expected due to variability in human input and practical system constraints that prevent ideal kinematic scaling. Moreover, standard deviations across trials per motion demonstrate consistency for both traditional and proposed methods. The average standard deviations are 0.107 for traditional and 0.057 for the proposed method, with maximum values not exceeding 0.285 and 0.146, respectively.

\begin{table*}[t]
\centering
\small
\setlength{\tabcolsep}{3pt}
\caption{Performance comparison: completion time and smoothness metrics. Bold indicates improvements (lower is better).}
\label{tab:results_smoothness}
\begin{tabular}{ll*{3}{ccc}}
\toprule
\multirow{2}{*}{\textbf{Motion}} & \multirow{2}{*}{\textbf{Index}}
& \multicolumn{3}{c}{\textbf{Completion Time (s)}}
& \multicolumn{3}{c}{$\boldsymbol{\Sigma}$ \textbf{Peak} $|d\tau/dt|$ \textbf{(Nm/s)}}
& \multicolumn{3}{c}{$\boldsymbol{\Sigma}$ \textbf{RMS} $d\tau/dt$ \textbf{(Nm/s)}} \\
\cmidrule(lr){3-5}\cmidrule(lr){6-8}\cmidrule(lr){9-11}
& & Trad. & Prop. & $\Delta$ (\%)
  & Trad. & Prop. & $\Delta$ (\%)
  & Trad. & Prop. & $\Delta$ (\%) \\
\midrule
Arbitrary & 1 & 7.77 & 7.85 & +0.9 & 994 & 832 & \textbf{-16.3} & 267.0 & 250.9 & \textbf{-6.0} \\
\midrule
\multirow{3}{*}{Ellipse}
& 1 & 5.82 & 5.81 & \textbf{-0.0} & 1048 & 956 & \textbf{-8.8} & 246.3 & 243.4 & \textbf{-1.2} \\
& 2 & 5.48 & 5.40 & \textbf{-1.3} & 865 & 867 & +0.3 & 223.4 & 243.4 & +8.9 \\
& 3 & 4.05 & 4.09 & +1.0 & 1156 & 908 & \textbf{-21.4} & 316.0 & 273.5 & \textbf{-13.4} \\
\midrule
\multirow{3}{*}{Infinity}
& 1 & 7.10 & 7.14 & +0.6 & 1131 & 1023 & \textbf{-9.5} & 278.2 & 270.9 & \textbf{-2.6} \\
& 2 & 5.79 & 5.88 & +1.6 & 1018 & 1083 & +6.4 & 282.9 & 287.4 & +1.6 \\
& 3 & 4.65 & 4.69 & +0.9 & 1013 & 942 & \textbf{-7.0} & 281.0 & 265.7 & \textbf{-5.4} \\
\midrule
\multirow{3}{*}{Sinusoidal}
& 1 & 7.80 & 7.77 & \textbf{-0.4} & 1053 & 1072 & +1.9 & 304.4 & 268.6 & \textbf{-11.8} \\
& 2 & 5.49 & 5.48 & \textbf{-0.2} & 1027 & 1026 & \textbf{-0.1} & 297.3 & 285.3 & \textbf{-4.0} \\
& 3 & 6.08 & 6.06 & \textbf{-0.3} & 1139 & 905 & \textbf{-20.5} & 289.2 & 260.5 & \textbf{-9.9} \\
\bottomrule
\end{tabular}
\end{table*}

\subsubsection{Quantitative Performance Results}
Table~\ref{tab:results_smoothness} presents completion time and smoothness metrics (Section \ref{10}). Each row represents one motion pattern, and the reported values are the average of three runs for the same motion to account for stochasticity. The proposed method demonstrates considerable smoothness improvements in 7 of 10 motion patterns based on \textbf{Peak Torque Rate ($\Sigma$ Peak $|d\tau/dt|$)}. The maximum improvement reached -21.4\% (Ellipse Index 3), while the worst-case degradation was only +6.4\% (Infinity Index 2)—significantly smaller in magnitude. \textbf{RMS Torque Rate ($\Sigma$ RMS $d\tau/dt$)} shows even better performance with 8 of 10 patterns improved, including -13.4\%, -11.8\%, and -9.9\% reductions. 

Completion times remained within $\pm$2\% for all patterns, which is expected since, under the two-thirds power law, velocity decreases during sharp turns but increases along smoother segments. These compensatory effects balance overall task duration, confirming that smoothness improvements were achieved without compromising task efficiency.

\section{CONCLUSIONS}
In this paper, we presented a shared control teleoperation framework for robotic arm end effectors. The primary objective was to generate human like output motions, motivated by their importance in Human Robot Interaction and their influence on overall system performance. The proposed architecture establishes shared control between the operator and the robotic system, whereby the controller modulates the user intended commands to embed human hand kinematic characteristics governed by the two thirds power law. The framework was experimentally validated on a Dobot CR10 robotic arm. Results demonstrated that the executed trajectories exhibit stronger adherence to human hand kinematics compared to conventional teleoperation. These improvements were further reflected in enhanced motion smoothness, as quantified by torque based metrics. Future work will explore additional biological motion principles beyond the two-thirds power law, as the proposed formulation is expected to generalize to other human-inspired kinematic models. In particular, we will incorporate models capable of handling 3D motions, as the present study considered only planar motions. Moreover, we aim to evaluate the proposed framework through subjective assessment, as we focused in this primarily on quantitative evaluation.




\bibliographystyle{IEEEtran}
\bibliography{references}

\end{document}